\documentclass[a4paper,12pt]{article}
\usepackage{authblk}

\usepackage{url}
\usepackage{hyperref}
\usepackage{amssymb}
\usepackage{amsmath}
\usepackage{array}
\usepackage{graphicx}
\usepackage{xcolor}
\usepackage{multirow}

\newcommand{\apj}{Astrophysical Journal}

\usepackage{subfigure}

\begin{document}

\title{\large\bf Singularity Resolution in Quantum Cosmology via
Page-Wootters Formalism}
 
\author{\normalsize\bf Vishal\thanks{\tt vishal.phiitg@iitg.ac.in} \ and Malay K. Nandy\thanks{\tt mknandy@iitg.ac.in \rm (Corresponding Author)} \\ \normalsize\em Department of Physics, Indian Institute of Technology Guwahati \\ \normalsize\em Guwahati 781 039, India} 

          
\date{\normalsize\rm (7 May 2026)}

\maketitle  
\begin{abstract}
We investigate the problem of classical big bang singularity in a plane-symmetric Bianchi type-I universe within the Wheeler-DeWitt (WDW) framework of quantum gravity. To address the problem of time, we employ the Page-Wootters formalism, which provides a relational notion of dynamics by conditioning the global state on a clock subsystem. Using Misner variables, the WDW equation assumes a Klein-Gordon (KG) type form. Its general solution is constructed as a Gaussian superposition of momentum eigenstates, resulting in an entangled global state between the clock and the remaining subsystem. Within this relational framework, we construct conditional states and obtain the corresponding probability density consistent with the KG-type inner product. The resulting conditional probability density vanishes in the limit of zero volume for all clock values, indicating quantum resolution of the classical singularity. We further show that positivity of the probability density imposes constraints on the admissible clock values,  which depend on the parameters of the Gaussian wavepacket. These results highlight the essential  role of quantum correlations in the emergence of relational dynamics, and demonstrate  that the Page-Wootters formalism provides a consistent and nonsingular probabilistic description of quantum cosmology.\\

\end{abstract}

\tableofcontents

\section{Introduction}
\label{introduction}
The concept of time is of fundamental importance in all areas of physics as no meaningful description of dynamics can be formulated without time. However, the precise role and interpretation of time have undergone significant conceptual evolution across different physical theories~\cite{smolin2013time}. 

In the Newtonian view, time is an external and absolute parameter, independent of the system and unaffected by dynamics, providing a universal background for evolution. This picture is fundamentally altered in general relativity, where time is no longer absolute but becomes intertwined with space as part of a four-dimensional spacetime manifold. In this framework, time is dynamical and observer-dependent, shaped by the gravitational field~\cite{landau1971classical}. In contrast, quantum mechanics retains a notion of time closer to the classical one: time enters as an external parameter governing the unitary evolution of quantum states, rather than as an operator on the same footing as other observables~\cite{dirac1981principles}.

Despite these differences, both classical and quantum theories exhibit time-reversal symmetric fundamental laws, while at the macroscopic level we observe an arrow of time, most prominently associated with the growth of entropy~\cite{hawkingarrow}. This contrast between time-symmetric laws and time-asymmetric phenomena highlights the intricate nature of time in physical theories.

The situation becomes even more subtle when attempting to reconcile general relativity with quantum mechanics. In canonical approaches to quantum gravity, such as the Wheeler–DeWitt(WDW) framework, the notion of time disappears from the fundamental equations~\cite{WheelerQG,dewitt1967quantum}. The Wheeler–DeWitt equation is a constraint equation with no explicit time dependence, thereby giving rise to a static or \textit{frozen} wavefunction of the universe. The resulting \textit{timeless} description poses a conceptual challenge: how can one recover the notion of dynamical evolution and temporal ordering from a theory in which time is not explicitly present? This is commonly referred to as the problem of time in quantum  gravity~\cite{kuchavr2011time,isham1993canonical}.

This issue has received considerable attention since the early formulations of quantum gravity~\cite{wald1980quantum,unruh1989time,teitelboim1982,Julian1994}. The problem of time comprises several difficulties that arise in attempting to define time in quantum gravity. This includes the absence of a preferred time variable, challenges in defining observables, and ambiguities in constructing the physical Hilbert space~\cite{anderson2012problem}. Another question arises in this context: how and when a notion of time should be identified---whether prior to quantization, through a classical choice of time variable, or only after quantization as an emergent concept. Closely related to this is the challenge of defining a consistent probabilistic interpretation, since the notion of probability conservation in quantum theory relies on a well-defined time parameter governing the evolution~\cite{isham1993canonical}.

The problem of time becomes especially pronounced in quantum cosmology, where the universe is treated as a closed system without any external observer or external time parameter. In this setting, the absence of a preferred notion of time makes the interpretation of quantum dynamics highly nontrivial. As widely discussed in the literature, this limitation restricts the applicability of the canonical framework in addressing key issues such as early-universe evolution and the resolution of the cosmological singularity~\cite{feinberg1988,kiefer_zeh1995,Ashtekar_2011}. Consequently, the problem of time in quantum cosmology is not merely a formal issue but is directly tied to the physical interpretation of cosmic evolution.

These issues suggest that time cannot be introduced as an external parameter in quantum cosmology, but must instead be identified intrinsically from within the dynamical degrees of freedom of the system~\cite{rovelli1991time}. In this context, several approaches have been developed in which evolution is formulated without reference to a fundamental time variable. One such perspective is the relational viewpoint, in which physical predictions are expressed in terms of correlations between observables, so that the change of one set of variables is described with respect to another that effectively plays the role of a clock~\cite{rovelliQM,hohn2021trinity}. Closely related ideas also appear in timeless formulations of quantum theory, such as the consistent histories approach, where probabilities are assigned to the entire history rather than to states evolving in time~\cite{griffiths1984consistent,gell1996quantum,hartle2014spacetime}. These frameworks indicate that a notion of dynamics can be recovered through correlations and histories, providing a way to address both the absence of an external time parameter and the formulation of a consistent probabilistic interpretation.

Among the various approaches that implement this relational notion of dynamics, the Page–Wootters formalism provides a particularly clear and concrete realization within the framework of quantum gravity~\cite{page1983evolution}. In this approach, the total system is described by a global stationary quantum state satisfying a constraint equation, such as the Wheeler–DeWitt equation, $\hat{H}|\Psi\rangle = 0$.

The central idea is that, for a closed system, an external time parameter is not directly observable, and it is therefore consistent to describe the global state as stationary. A notion of dynamics nevertheless emerges by partitioning the system into two subsystems, a clock and the remaining degrees of freedom, thereby defining the evolution relationally through conditional states with respect to the clock. In this way, the apparent temporal behavior of the system is understood as arising from correlations between subsystems, rather than from evolution with respect to an external time parameter. Under suitable conditions, this relational description reproduces an effective Schrödinger-type evolution for the remaining degrees of freedom. This makes the framework particularly suitable for quantum cosmology, where the absence of an external time parameter requires that the dynamics be understood in purely relational terms~\cite{Gambini_2004,gambini2009conditional,Giovannetti2015,mv2017}.

While the Page–Wootters framework provides a consistent way to recover relational dynamics in the absence of an external time parameter, a primary motivation for quantum cosmology is resolution of the classical singularity, where the spacetime description itself breaks down. The most notable example is the big bang singularity, in which the universe originates from a state of vanishing volume, prior to which classical spacetime cannot be extended.

For investigating the dynamics of the early universe within quantum cosmology, the most widely studied cosmological minisuperspace models are the Friedmann–Lemaître–Robertson–Walker (FLRW)~\cite{friedman1922krummung,lemaitre1927univers,robertson1935kinematics,walker1937milne} and Bianchi~\cite{ryan2015homogeneous,ellis2006bianchi} spacetimes. The FLRW model describes a perfectly homogeneous and isotropic universe, whereas Bianchi models generalize this assumption by relaxing isotropy while maintaining spatial homogeneity.

The WDW equation has been widely applied in the study of cosmological singularity using these models of the universe. Early investigations indicated that imposing appropriate boundary conditions--such as requiring the wavefunction to vanish at zero spatial volume could effectively avoid the classical big bang singularity and quantum effects may resolve or significantly modify the classical singular behavior~\cite{dewitt1967quantum,misner1969quantum,hartle1983wave}. Subsequent works with these ideas suggested that quantum effects within the WDW framework can modify the classical evolution, leading to scenarios in which the singularity is avoided and replaced by features such as quantum bounces or oscillatory behavior in both FLRW and Bianchi spacetimes~\cite{kiefer2019singularity,bouhmadi2014resolution}.

An alternative and extensively studied approach to singularity resolution is provided by loop quantum cosmology (LQC), which arises from the application of loop quantum gravity(LQG) techniques to symmetry-reduced cosmological models. In this framework, the classical continuum description of spacetime is replaced by a discrete quantum geometry, leading to significant modifications of the early-universe dynamics and singularity avoidance~\cite{bojowald2001,ashtekar2006}.

Despite significant progress in addressing the problem of time and the resolution of cosmological singularity by quantum gravity, a framework that consistently treats both issues remains lacking. While the Wheeler–DeWitt approach provides insights into singularity avoidance, it suffers from the absence of a well-defined notion of time, complicating the interpretation of dynamics. In contrast, the Page–Wootters formalism offers a relational description of evolution without an external time parameter, but its implications for singularity resolution have not been thoroughly explored. 

This motivates the present work, in which we investigate the big bang singularity within the Page–Wootters framework. We consider a plane-symmetric Bianchi type-I universe formulated in a relational setting, where evolution is defined with respect to an internal clock. By partitioning the system into clock and rest subsytems, we construct conditional states and probabilities to define evolution with respect to the internal clock. We find that the probability of vanishing volume is zero, indicating a nonsingular quantum description of the early universe.

The rest of the paper is organized as follows. In Section~\ref{pw-framework}, we review the Page–Wootters framework. Section~\ref{entanglement} presents solution of the Wheeler–DeWitt equation and the construction of entangled states. In Section~\ref{conditional-universe}, we analyze the conditional state and the associated Klein-Gordon probability. Finally, we conclude the paper in  Section~\ref{discussion} with a discussion.

\section{The Page--Wootters Formalism}
\label{pw-framework}
The Page--Wootters formalism~\cite{page1983evolution} provides a rigorous framework for understanding relational dynamics in quantum systems without an external time parameter. It addresses the problem of time in quantum gravity by treating time as an internal degree of freedom, a quantum clock subsystem within a globally timeless state of the universe~\cite{emergence2023,rijavec2025}. Since there is no external time parameter, the physical states satisfy the Wheeler--DeWitt constraint,
\begin{equation}
\hat{H}|\Psi\rangle = 0.
\label{wd}
\end{equation}
This expresses global stationarity of the universe, encapsulating the timeless nature of the total quantum state. To obtain a relational dynamics by choosing a clock from the closed system, the total Hilbert space of the universe $\mathcal{H} $ factorizes into two subsystems: the clock and the rest of the universe,  
\begin{equation}
\mathcal{H} = \mathcal{H}_C \otimes \mathcal{H}_R,
\end{equation}
where $\mathcal{H}_C$ is the Hilbert space associated with the clock subsystem, and $\mathcal{H}_R$ represents the remaining part of the universe whose evolution we want to describe relationally. The condition for a feasible clock is that it should not interact with the rest of the universe so that the Hamiltonian can be written as,
\begin{equation}
\hat{H} = \hat{H}_C \otimes \mathbb{I}_R + \mathbb{I}_C \otimes \hat{H}_R,
\label{hamiltonian}
\end{equation}
where $\hat{H}_C$ and $\hat{H}_R$ act only on the clock and rest Hilbert spaces, respectively. 

The clock is characterized by an observable $T_C$ conjugate to $H_C$, $[T_C, H_C] = i$, with eigenstates $| t \rangle_C$ satisfying $T_C | t \rangle_C = t | t \rangle_C$, where $t \in \mathbb{R}$ labels clock hands. Now, in order to set the dynamics of the rest of the universe with respect to the clock states, the entanglement between quantum states of clock and rest plays a crucial role~\cite{mv2017,rijavec2023}. The state of the universe can be expanded as
\begin{equation}
| \Psi \rangle = \sum_t c_t | t \rangle_C \, |\psi_t \rangle_R,
\label{eq:entangled}
\end{equation}
with multiple nonzero $c_t$, encoding the appearance of dynamical evolution of the universe. Here the symbol $\sum_t \ldots$ is interpreted as $\int dt \ldots$. Henceforth, we shall denote $|t \rangle_C$ by  
$|t \rangle$ for brevity.

The total density operator is expressed as $\rho = |\Psi\rangle \langle \Psi|$ and the conditional state of the rest subsystem at clock reading $t$ is given by Everett's relative state~\cite{evrett1957},
\begin{equation}
\rho_R (t) = \frac{ \mathrm{Tr}_C \left[ (|t\rangle\langle t| \otimes \mathbb{I}_R) \rho \right]}{\mathrm{Tr} \left[ (|t\rangle\langle t| \otimes \mathbb{I}_R) \rho\right]} = \  | \psi_t \rangle_R \, _R\langle \psi_t|,
\label{eq:relative}
\end{equation}
where no measurement on the clock occurs. The clock state evolves as $|t\rangle = e^{-i \hat{H}_C t} |0\rangle$ for some clock eigenstate $|0 \rangle$ as the intial time. Now if we use the decomposed form of the total Hamiltonian of the system~\eqref{hamiltonian} in the Hamiltonian constraint~\eqref{wd}, we obtain
\begin{equation}
\hat{H}|\Psi\rangle = \sum_t c_t \left(
\hat{H}_C |t\rangle \otimes |\psi_t\rangle_R
+
|t\rangle \otimes \hat{H}_R |\psi_t\rangle_R
\right)
= 0 .
\label{heq}
\end{equation}
Using $\hat{H}_C |t\rangle = i \frac{\partial}{\partial t} |t\rangle$ and projecting onto a clock state $\langle t'|$, we obtain from Eq.~\eqref{heq} and Eq.~\eqref{wd}
\begin{equation}
\langle t'| \hat{H}|\Psi\rangle=\sum_t c_t \left(
i \frac{\partial \, \langle t'|t \rangle }{\partial t} |\psi_t\rangle_R
+ \, \langle t'|t\rangle \hat{H}_R |\psi_t\rangle_R
\right)
= 0 .
\label{heq2}
\end{equation}
Now using the orthogonality condition for the clock states, $\langle t'|t\rangle = \delta(t' - t)$, and upon integration by parts, Eq.~\eqref{heq2} simplifies to the Schrodinger equation for the state of the remaining subsystem,
\begin{equation}
i \frac{\partial}{\partial t} |\psi_t\rangle_R = \hat{H}_R |\psi_t\rangle_R .
\end{equation}

Thus, the standard Schrodinger evolution for the subsystem emerges from the global stationary state implied by the WDW equation, $\hat{H}|\Psi\rangle = 0$, together with the quantum evolution of the clock states given by $\hat{H}_C |t\rangle = t |t \rangle$. The dynamics emerges unitarily on the subsystem $R$ with respect to the clock variable $t$, from entanglement between $C$ and $R$.

Given the conditional state of the subsystem $R$ at clock reading $t$, we can define physical predictions for the subsystem conditioned on the clock reading $t$. Let $|\phi \rangle$ denote an eigenstate of an observable $\hat{O}$ in the subsystem $R$. The conditional probability of obtaining the value $\phi $ when the clock reads $t$ is given by
\begin{equation}
P(\langle \phi|t \rangle) = \frac{\mathrm{Tr} \left[ (|t\rangle\langle t| \otimes |\phi \rangle\langle \phi|) \rho \right]}{\mathrm{Tr} \left[ (|t\rangle\langle t| \otimes \mathbb{I}_R) \rho \right]}.
\label{conditional-probability}
\end{equation}
For the observable $\hat{O}$ acting on the subsystem $R$, the conditional expectation value at clock reading $t$ is defined as
\begin{equation}
\langle \hat{O} \rangle_t =\frac{\mathrm{Tr} \left[ (|t\rangle\langle t| \otimes \hat{O}) \rho \right] }{\mathrm{Tr} \left[ (|t\rangle\langle t| \otimes \mathbb{I}_R) \rho\right]}.
\label{expectation}
\end{equation}

These expressions show that all physical predictions in the Page--Wootters framework are obtained conditionally, with the clock degree of freedom providing a reference against which the rest of the system evolves. Even though the total state $|\Psi\rangle$ is stationary, the conditional state $|\psi_t\rangle_R$ exhibits non-trivial dynamics, reproducing the usual quantum mechanical evolution in a relational manner.

\section{Bianchi Type-I universe and clock variable}
\label{entanglement}
Bianchi universes \cite{landau1971classical,ellis1999cosmological,misner1972magic} are the simplest (classical) anisotropic models which are homogenous but do not respect spatial isotropy. A particularly interesting subclass of Bianchi Type-I models is the plane symmetric or ellipsoidal universe, where two of the three spatial directions expand (or contract) identically, while the third evolves independently. The line element for the plane symmetric Bianchi Type-I universe is given by
\begin{equation}
ds^2 = - N dt^2  +a^2(t) (dx^2 +  dy^2) + b^2(t) dz^2
\end{equation}
where $a(t)$ and $b(t)$ are scale factors along the three different axes. We consider this anisotropic model in the minisuperspace representation of the Wheeler-DeWitt framework of quantum gravity \cite{WheelerQG,dewitt1967quantum} to study the behavior of the universe in the early stage of evolution. 
where we have chosen the $xy$ plane as the plane of symmetry. The determinant of the metric for this model is $\sqrt{h} = a^2 b$ and extrinsic curvature scalar $K   = \frac{-1}{N} (\frac{2 \dot{a}}{a} +\frac{\dot{b}}{b} )$ and Ricci scalar ${}^3\!R = 0$.
Now we transform variables, $\alpha = \log( a^{2} b)$ and $\beta = \frac{1}{2} \log(\frac{b^2}{a^2})$. We will treat $\alpha$ as volume and $\beta$ as anisotropy parameter. The action can be written as
\begin{equation}
S = \frac{1}{4 \pi G} \int dt \, d^3x  \frac{e^{3 \alpha}}{N}  \left( \dot{\beta}^2-\dot{\alpha}^2 \right).
\end{equation}
The total Hamiltoniam of the system is written as
\begin{equation}
H = p^{2}_{\beta} - p^{2}_{\alpha},
\end{equation}
where $p_\alpha$ and $p_\beta$ are momenta conjugate to the variables $\alpha$ and $\beta$, respectively. We now quantize this minisuperspace model canonically by promoting the classical momenta to quantum operators, $p_{\beta} \rightarrow -i\hbar\frac{\partial}{\partial \beta},\quad p_{\alpha} \rightarrow -i\hbar\frac{\partial}{\partial \alpha}$, and by substituting into the Hamiltonian constraint $H=0$. This yields the Wheeler-DeWitt equation,
\begin{equation}
\left[
\frac{\partial^2}{\partial \alpha^2}
-\frac{\partial^2}{\partial \beta^2}
\right] \Psi(\alpha, \beta) = 0.
\label{wdeq}
\end{equation}

To apply the Page--Wootters relational formalism, we choose the
anisotropy variable $\beta$ as the clock degree of freedom and the volume variable $\alpha$ as the remaining subsystem. The total Hamiltonian operator factorizes as
\begin{equation}
\hat{H} = \hat{H}_C \otimes \mathbb{I}_R + \mathbb{I}_C \otimes \hat{H}_R,
\end{equation}
where $\hat{H}_C = \hat{p}_\beta^2 $ and $\hat{H}_R = -\hat{p}_\alpha^2$ are the Hamiltonian of the clock and rest of the universe, respectively. In contrast to the standard PW construction, where the clock Hamiltonian is linear in the conjugate momentum, the present model involves the term quadratic in the momentum operator. As a result, the clock Hamiltonian does not generate translations in the clock variable.

The quadratic nature of the Hamiltonian constraint leads to a Klein–Gordon-type equation, in which no preferred time evolution exists at the fundamental level. Instead of imposing a frequency restriction, we retain the full structure and interpret the clock variable relationally within the PW framework. The clock variable is defined as canonically conjugate to the momentum operator, satisfying the commutation relation $[\hat{\beta}, \hat p_{\beta}] = i$.

In order to obtain the solution of the equation~\eqref{wdeq} we employ separation of variables and write the wavefunction as a superposition of left-moving and right-moving components as
\begin{equation}
\Psi(\alpha,\beta) = \int_{-\infty}^{\infty}  F(k) \, e^{i k \beta} \left( e^{i k \alpha} + e^{-i k \alpha} \right) dk,
\label{wffk}
\end{equation}
where $F(k)$ is the  mode distribution function and $k$ is the separation constant. This structure encodes a quantum superposition of expanding $e^{i k \alpha}$ and contracting $e^{-i k \alpha}$ branches of the quantum universe, while $e^{i k \beta}$ governs anisotropic evolution.

We thus consider the total state written in the momentum basis $|k \rangle$ as
\begin{equation}
|\Psi\rangle = \int dk \ F(k) \ |k\rangle_C \otimes |\psi_k\rangle_R ,
\label{entgaled}
\end{equation}
with  $F(k)$ determining the superposition of momentum modes that encodes the correlation between the clock and the remaining subsystem. This form~\eqref{entgaled} makes the entangled structure of the state $|\Psi\rangle$ explicit, with each clock momentum mode $|k\rangle_C$ correlated with a corresponding state $|\psi_k\rangle_R$ of the remaining subsystem.

The momentum eigenstates in the clock and the subsystem sectors are defined through 
\begin{equation}
\langle \beta | k \rangle_C = \frac{1}{\sqrt{2\pi}} e^{i k \beta},
\qquad
\langle \alpha | k \rangle_R = \frac{1}{\sqrt{2\pi}} e^{\pm i k \alpha}.
\label{plane-structure}
\end{equation}
The plane-wave structure in~\eqref{plane-structure} reflects that $|k\rangle_{C,R}$ are eigenstates of the corresponding momentum operators $\hat{p}_\beta$ and $\hat{p}_\alpha$, respectively. The action of the corresponding Hamiltonians on these momentum eigenstates lead to
\begin{equation}
\hat{H}_C |k\rangle_C = k^2 |k\rangle_C,
\qquad
\hat{H}_R |k\rangle_R = -k^2 |k\rangle_R,
\end{equation}
which follow directly from $\hat{H}_C = \hat{p}_\beta^2 $ and $\hat{H}_R = -\hat{p}_\alpha^2$. The opposite signs reflect the hyperbolic (Klein--Gordon-type) structure of the equation, ensuring that the total Hamiltonian constraint $\hat{H}_C + \hat{H}_R = 0$ is satisfied for each mode.

The clock parameter $\beta$ is conjugate to the momentum operator $\hat{p}_{\beta}$ (since $[\beta , \hat{p}_{\beta} ] =i $) and forms a continuous labeling of the clock readings. The eigenstates $|\beta\rangle$ of the clock Hamiltonian satisfies the orthonormality condition
\begin{equation}
\langle \beta | \beta' \rangle =\int dk \langle \beta | k \rangle_C \ {}_C\langle k | \beta' \rangle =\frac{1}{2\pi} \int dk e^{ik(\beta-\beta')} = \delta(\beta - \beta'),
\end{equation}
and therefore forms an orthogonal basis in the clock Hilbert space. This shows that $| \beta \rangle$ represents distinguishable clock states. Furthermore, the vanishing commutator $[\hat{H}_C , \hat{H}_R ]=0$ indicates that the clock does not interact with the remaining subsystem. In addition, the total global state $|\Psi \rangle$ represents an etangled structure between the clock and the remaining susbsytem. 

The above properties ensure that $\beta$ provides a valid clock variable in the Page--Wootters construction. In particular, the orthogonality guarantees distinguishable clock readings, while the absence of direct interaction with the remaining subsystem preserves its role as a reference degree of freedom. In addition, the entangled structure of the total state establishes correlations between the clock and the remaining subsystem, which is essential for defining conditional states and for generating relational dynamics~\cite{Giovannetti2015,mv2017,moreva2014}.

\section{Conditional states and Klein-Gordon probability}
\label{conditional-universe}
We construct the total wavefunction by superposing momentum eigenstates with a Gaussian weight
\[
F(k) = \sqrt{\frac{1}{\sigma}\sqrt{\frac{2}{\pi}}}\; e^{-\frac{(k - k_0)^2}{\sigma^2}},
\]
which is normalized such that $\int dk\, |F(k)|^2 = 1$. The total wavefunction then follows from Eq.~\eqref{wffk} upon carrying out the integration over $k$, leading to
\begin{equation}
\Psi(\alpha,\beta)=  \sqrt{ \sigma \pi} \left(\frac{2} {\pi} \right)^{\frac{1}{4}}
\left[
e^{-\frac{\sigma^2}{4}(\alpha+\beta)^2} e^{i k_0 (\alpha+\beta)}
+
e^{-\frac{\sigma^2}{4}(\alpha-\beta)^2} e^{-i k_0 (\alpha-\beta)}
\right].
\label{wffull}
\end{equation}
This state represents a superposition of two Gaussian wavepackets peaked around $k = \pm k_0$, propagating in opposite directions in the minisuperspace. 

We normalize the wavefunction using the Klein--Gordon inner product. Since \(\beta\) is identified as the time variable, the inner product is defined as
\begin{equation}
\langle \Psi | \Psi \rangle_{KG} = i \int_{-\infty}^{\infty} d\alpha \left[ 
\Psi^*(\alpha,\beta)\,\frac{\partial \Psi(\alpha,\beta)}{\partial \beta}
-
\frac{\partial \Psi^*(\alpha,\beta)}{\partial \beta}\,\Psi(\alpha,\beta)
\right].
\label{KG-inner-product}
\end{equation}
Using the wavefunction~\eqref{wffull} in the above expression, the Klein-Gordon inner product becomes
\begin{align}
\langle \Psi | \Psi \rangle_{KG} = \sigma \pi \sqrt{\frac{2} {\pi}} \int_{-\infty}^{\infty} d\alpha \
\Bigg[ & \ e^{-\frac{\sigma^2}{2}(\alpha^2+\beta_0^2)} \Big\{4 k_0 \cosh(\sigma^2 \alpha \beta_0) \nonumber \\ 
& + 4 k_0 \cos(2 k_0 \alpha)- 2 \sigma^2 \alpha \sin(2 k_0 \alpha) \Big\} \Bigg].
\label{kg-integral}
\end{align}
Integrating over $\alpha$, Eq.~\eqref{kg-integral} yields
\begin{equation}
\langle \Psi | \Psi \rangle_{KG} = 8 \pi k_0.
\label{inner-product}
\end{equation}
Although the terms involving the cosine and sine functions give non-zero contributions, they cancel exactly because they yield contributions of equal magnitude  with opposite signs.

For the Klein--Gordon inner product in Eq.~\eqref{inner-product} to be positive and non-zero, it is necessary that \(k_0 > 0\). This necessity stems from the well-known feature that the Klein--Gordon inner product is not positive definite in general, and that positivity must be enforced by an appropriate choice of states. 

Using the normalization obtained from the Klein--Gordon inner product~\eqref{inner-product}, we write the normalized wavefunction in the form
\begin{equation}
\Psi_N(\alpha,\beta)= N \left[
e^{-\frac{\sigma^2}{4}(\alpha+\beta)^2} e^{i k_0 (\alpha+\beta)}
+
e^{-\frac{\sigma^2}{4}(\alpha-\beta)^2} e^{-i k_0 (\alpha-\beta)}
\right],
\label{norm-wf}
\end{equation}
where the normalization constant is given by
\[
N = \sqrt{\frac{\sigma}{8 k_0}}\,\left(\frac{2}{ \pi}\right)^{\frac{1}{4}}.
\]
Having identified \(\beta\) as the clock variable, we now proceed to construct the conditional state corresponding to a definite clock reading \(\beta = \beta_0\). This is achieved by projecting the total wavefunction onto the clock eigenstate \(|\beta_0\rangle\), following the Page--Wootters prescription.
The total state can be expressed as
\begin{equation}
|\Psi_N \rangle = \int d\alpha \, d\beta \ \Psi_N(\alpha,\beta)\, |\beta\rangle \otimes |\alpha\rangle_R .
\end{equation}

To extract the relational dynamics, we project the total state onto a definite clock state $|\beta_0\rangle$. This is implemented by applying the projection operator $|\beta_0\rangle \langle \beta_0| \otimes \mathbb{I}_R$, yielding
\begin{equation}
|\beta_0\rangle \langle \beta_0| \otimes \mathbb{I}_R |\Psi_N \rangle 
= \int d\alpha \, d\beta \ \Psi_N (\alpha,\beta)\, |\beta_0\rangle \langle \beta_0 | \beta \rangle \otimes |\alpha\rangle_R.
\end{equation}

Using the orthonormality condition $\langle \beta_0 | \beta \rangle = \delta(\beta - \beta_0)$, the integration over $\beta$ can be performed explicitly, leading to
\begin{equation}
|\beta_0\rangle \langle \beta_0| \otimes \mathbb{I}_R |\Psi_N\rangle 
= |\beta_0\rangle \otimes \int d\alpha \ \Psi_N(\alpha,\beta_0)\, |\alpha\rangle_R.
\label{projection}
\end{equation}

We thus identify the conditional state of the remaining subsystem at clock time $\beta_0$ as
\begin{equation}
|\psi_{\beta_0}\rangle_R = \langle \beta_0| \otimes \mathbb{I}_R |\Psi_N\rangle  = \int d\alpha \ \Psi_N(\alpha,\beta_0)\, |\alpha\rangle_R .
\label{condtn}
\end{equation}

The corresponding conditional wavefunction is therefore given by
\begin{equation}
\Psi_N (\alpha,\beta_0) = N \left[ e^{-\frac{\sigma^2}{4}(\alpha+\beta_0)^2} e^{i k_0 (\alpha+\beta_0)} + e^{-\frac{\sigma^2}{4}(\alpha-\beta_0)^2} e^{-i k_0 (\alpha-\beta_0)} \right].
\label{condnwf}
\end{equation}

This conditional state encodes the state of the remaining subsystem when the clock variable takes the value \(\beta_0\), and forms the basis for defining relational dynamics in the Page--Wootters framework. At first sight, the dependence on $\beta_0$ may appear trivial, as the conditional wavefunction in Eq.~\eqref{condnwf} is obtained by simply evaluating the total wavefunction at $\beta = \beta_0$. However, this projection has a clear physical interpretation: it corresponds to conditioning the global state on a definite reading of the clock, thereby selecting a relational slice of the total state $| \Psi_N \rangle$, as implied by Eq.~\eqref{projection}.

The nontrivial content of this construction lies in the entanglement structure of the total wavefunction. The clock and the remaining degree of freedom are correlated, and it is precisely these correlations that give rise to an effective notion of dynamics. With different choices for \(\beta_0\), the conditional wavefunction $\Psi_N (\alpha,\beta_0)$ of the remaining subsystem is different, reflecting how different configurations of the subsystem are correlated with different readings of the clock.

In this sense, the parameter \(\beta_0\) does not represent an external time variable, but rather labels different conditional states of the remaining subsystem relative to the clock. Having obtained the conditional state, we can now define the corresponding conditional probability density for the variable \(\alpha\), which allows us to extract physically meaningful predictions within this relational framework.

The conditional probability density $P(\alpha|\beta_0)$ with respect to the volume parameter $\alpha$ is defined as
\begin{equation}
P(\alpha | \beta_0) = i \left(  \Psi_N^{*}(\alpha,\beta_0) \frac{\partial \Psi_N(\alpha,\beta_0)}{\partial \beta_0} - \frac{\partial \Psi_N^{*}(\alpha,\beta_0)}{\partial \beta_0} \Psi_N(\alpha,\beta_0) \right),
\end{equation}
ensuring consistency with the structure of inner product~\eqref{KG-inner-product}. Using the conditional wavefunction $\Psi_N (\alpha,\beta_0)$ from Eq.~\eqref{condnwf}, we obtain the conditional probability density for the variable $\alpha$ at a fixed clock reading $\beta_0$ as
\begin{align}
P(\alpha | \beta_0) = \frac{\sigma}{8 } \sqrt{\frac{2}{\pi}} \ e^{-\frac{\sigma^2}{2}(\alpha^2+\beta_0^2)}
\Big[ & 2 \cosh(\sigma^2 \alpha \beta_0) + 2 \cos(2 k_0 \alpha)- \frac{ \sigma^2 \alpha}{k_0} \sin(2 k_0 \alpha) \Big].
\label{probability}
\end{align}
It is evident from the above expression that the conditional probability density vanishes in the limit $\alpha \to -\infty$,
\begin{equation}
\lim_{\alpha \to -\infty } P(\alpha| \beta_0) = 0,
\end{equation}
independent of $\beta_0$. Independence of this limit with respect to $\beta_0$ implies that the probability of obtaining zero volume vanishes irrespective of how we set the clock reading $\beta_0$.

This result implies that configurations corresponding to vanishing volume, $\alpha \to -\infty$, are suppressed at the level of conditional probabilities for all clock values. Since this limit corresponds to  the classical big bang singularity, the model exhibits a clear resolution of the classical singularity within the Page-Wootters framework. In this sense, the emergence of relational dynamics, driven by the correlation between the clock and the remaining subsystem, leads to a nonsingular description of the quantum universe.

For finite values of the volume parameter $\alpha$, we must constraint the clock parameter $\beta_0$ so that the probability density remains positive definite. In order to carry out a proper analysis of the parameters we have to contrain the function 
\begin{equation}
F(\alpha ; \beta_0, k_0, \sigma) = \Big[ 2  \cosh(\sigma^2 \alpha \beta_0) + 2  \cos(2 k_0 \alpha)-\frac{\sigma^2 \alpha}{k_0}   \sin(2 k_0 \alpha) \Big]
\label{F-alpha}
\end{equation}
such that it remains positive for appropriate choices of the parameters involved. The exponential factor in equation~\eqref{probability} remains always positive and therefore it can be excluded from this analysis. 

We can reduce the number of parameters from three to two by subsituting $k_0 \alpha = x$, $ \frac{\sigma^2}{k_0^{2}} = \lambda $ and $\beta_0 k_0 = \mu $, so that the above function~\eqref{F-alpha} reduces to 
\begin{equation}
f(x; \lambda, \mu) = 2 \cosh(\lambda \mu x) + 2 \cos(2 x)- \lambda x \sin(2 x) =0.
\label{f-x}
\end{equation}

Depending on the magnitudes of the parameters $\lambda$ and $\mu$, whenever this function changes sign, it must necessarily pass through zero. Consequently, the roots of $f(x; \lambda, \mu) =0 $ determine the transition points separating positive and negative regions of the probability density. Thus we must constrain the parameter space ($\lambda , \mu$) so that no roots of equation~\eqref{f-x} exist for any $x$.
  
By numerical analysis of equation~\eqref{f-x}, we find conditions on minimum values of $\mu$ for a range of $\lambda$-values for which the function $f(x; \lambda, \mu)$, and hence the probability density~\eqref{probability} or equivalently~\eqref{prob2} below, would remain positive definite. These values are displayed in Table~\ref{tab:table-beta}, that also shows the corresponding values of $\sigma$ and $\beta_{0, \rm min}$ for different values of $k_0$.

\begin{table}[]
\centering
\caption{\small \sf Lower bounds on the parameter $\mu$ for different choices of $\lambda$ (shown in the first two columns) for positive definiteness of the probability density~\eqref{probability}. The subsequent columns show the corresponding lower bounds on the clock value $\beta_0$ for several choices of the mean $k_0$ and the width $\sigma$ of the Gaussian distribution for positive definiteness of the  probability density~\eqref{prob2}, or equivalently~\eqref{probability}.}
\label{tab:table-beta}
\begin{tabular}{|c|c|cc|cc|cc|}
\hline
\multirow{2}{*}{$\lambda $} & \multirow{2}{*}{$\mu_{\rm min} $} & \multicolumn{2}{c|}{$ k_0 =0.1$}          & \multicolumn{2}{c|}{$ k_0 = 1$}            & \multicolumn{2}{c|}{$k_0 =  10 $}         \\ \cline{3-8} 
                            &                         & \multicolumn{1}{c|}{$\sigma$} & $\beta_{0, \rm min}$ & \multicolumn{1}{c|}{$\sigma$} & $\beta _{0 , \rm min}$ & \multicolumn{1}{c|}{$\sigma$} & $\beta_{0 , \rm min}$ \\ \hline
0.001                       & 0.5                     & \multicolumn{1}{c|}{0.00316}  & 5         & \multicolumn{1}{c|}{0.0316}   & 0.5        & \multicolumn{1}{c|}{0.316}    & 0.05      \\ \hline
0.01                        & 0.5                     & \multicolumn{1}{c|}{0.01}     & 5         & \multicolumn{1}{c|}{0.1}      & 0.5        & \multicolumn{1}{c|}{1}        & 0.05      \\ \hline
0.1                         & 0.5                     & \multicolumn{1}{c|}{0.0316}   & 5         & \multicolumn{1}{c|}{0.316}    & 0.5        & \multicolumn{1}{c|}{3.162}    & 0.05      \\ \hline
1                           & 0.48                    & \multicolumn{1}{c|}{0.1}      & 4.8       & \multicolumn{1}{c|}{1}        & 0.48       & \multicolumn{1}{c|}{10}       & 0.048     \\ \hline
10                          & 0.27                    & \multicolumn{1}{c|}{0.316}    & 2.7       & \multicolumn{1}{c|}{3.162}    & 0.27       & \multicolumn{1}{c|}{31.62}    & 0.027     \\ \hline
100                         & 0.1                     & \multicolumn{1}{c|}{1}        & 1         & \multicolumn{1}{c|}{10}       & 0.1        & \multicolumn{1}{c|}{100}      & 0.01      \\ \hline
\end{tabular}
\end{table}

\begin{figure}[h!]
    \begin{center}
    {\includegraphics[width=1 \textwidth]{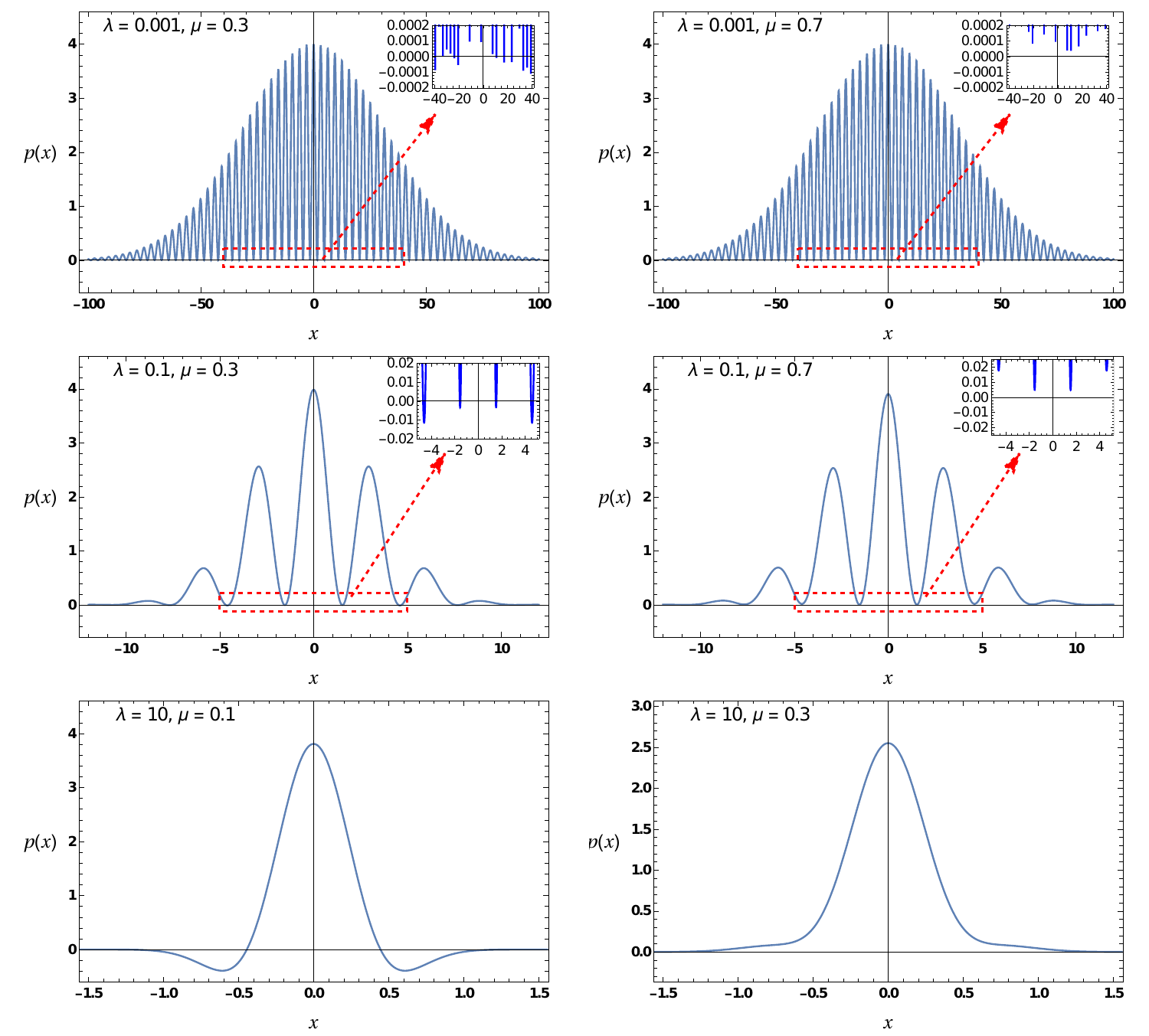}}\hfill%
     \end{center}
     \caption{\small \sf Conditional probability density $p(x;\lambda, \mu)$ given by~\eqref{prob2} for different values of the parameters $\lambda$ and $\mu$. Graphs in the left (right) column correspond to $\mu < \mu_{\rm min}$ ($\mu > \mu_{\rm min}$) (the lower bounds $\mu_{\rm min}$ for different values of $\lambda$ are shown in Table~\ref{tab:table-beta}). The insets illustrate the presence (absence) of negative probability regions for $\mu < \mu_{\rm min}$ ($\mu > \mu_{\rm min}$).}
    \label{fig:probab_plot}
\end{figure}

Figure~\ref{fig:probab_plot} shows the probability density~\eqref{probability} (leaving out the prefactors), written in the form,  
\begin{align}
p(x; \lambda, \mu) =\ e^{- \frac{\lambda}{2} (x^2+\mu^2)} \Big[2 \cosh(\lambda \mu x) + 2 \cos(2 x)- \lambda x \sin(2 x) \Big],
\label{prob2}
\end{align} 
illustrating that the probability density~\eqref{prob2}, and hence~\eqref{probability}, remains positive definite in accordance with parameter choices for $\lambda$ and $\mu$ displayed in Table~\ref{tab:table-beta}. Specifically, the figure illustrates that for the cases with $\mu > \mu_{\rm min}$, the probability density $p(x; \lambda, \mu)$ remains positive for any $x$, representing physical region of the parameter space ($\lambda, \mu$). On the other hand, for $\mu < \mu_{\rm min}$, there are regions in $x$ where the probability density $p(x; \lambda, \mu)$ becomes negative and hence represents unphysical region of the parameter space ($\lambda, \mu$).
  
Furthermore, it is evident from Table~\ref{tab:table-beta} that the mean value $k_0$ of the Gaussian distribution plays a crucial role in determining the lower bound on the clock value $\beta_0$.

When $k_0$ is near zero, Table~\ref{tab:table-beta} shows that lower bound on the clock value $\beta_0$ becomes high for $\sigma \ll k_0$ and the possibility of $\beta_{0, \rm min}$ close to zero can occur only when $\sigma \gg k_0$. On the other hand, for $k_0$ much away from zero, the lower bound on the clock value $\beta_0$ being close to zero becomes possible for both $\sigma \gg k_0$ and $\sigma \ll k_0$.

\section{Discussion and Conclusion}
\label{discussion}
In this work, we have investigated the issue of singularity in a plane-symmetric Bianchi type-I universe within the Wheeler--DeWitt framework of quantum gravity. To address the problem of time and extract a notion of evolution, we employed the Page--Wootters formalism, which provides a relational description of dynamics in a closed universe.

We describe the system in the minisuperspace constructed with Misner variables $\alpha$ and $\beta$. The Hamiltonian constraint yields the Wheeler-DeWitt equation $\hat{H} \Psi = (\hat{p}_\beta^2-\hat{p}_\alpha^2) \Psi =0 $. Choosing $\beta$ as the clock variable, satisfying $[\beta, \hat{p}_\beta] = i$, we decompose the total Hamiltonian into a clock and the remaining subsystem, $\hat{H}_C = \hat{p}_\beta^2$ and $\hat{H}_R = -\hat{p}_\alpha^2$, respectively. This decomposition is consistent with the Page-Wootters formalism as the clock does not interact with the remaining subsystem, reflected by $[\hat{H}_C , \hat{H}_R]=0$, and the clock eigenstates $|\beta \rangle$ form an orthogonal basis, reflected by $\langle \beta | \beta' \rangle = \delta(\beta - \beta')$. 

The Misner variables give a Klein-Gordon-type structure of the Wheeler--DeWitt equation, and the general solution of the Wheeler-DeWitt equation is constructed as a superposition of momentum eigenstates, weighted by a Gaussian distribution function $F(k)$, centered around $k_0$ with width $\sigma$. This function determines the spectral profile of the total state and encodes a correlation between the clock and the remaining subsystem. In particular, the global state $|\Psi \rangle$ can be viewed as an entangled superposition of clock momentum states $|k\rangle_C$ and the states of the remaining subsystem, $|\psi_k\rangle_R$.

It is precisely this entanglement structure that underlies the emergence of relational dynamics: conditioning on a definite clock reading $\beta_0$ selects the corresponding state of the remaining subsystem, thereby providing an effective notion of relational dynamics within a timeless framework. With the global state in hand, we construct the conditional wavefunction corresponding to the definite clock reading $\beta_0$ by projecting the total state onto the clock eigenstate $|\beta_0\rangle$.

With the total wavefunction so constructed, we normalized it in the framework of Klein--Gordon inner product, that required $k_0 > 0$ for the norm to be non-zero and positive definite. This procedure yields the state of the remaining subsystem conditioned on the clock value $\beta_0$, and provides the starting point for defining relational dynamics within the Page--Wootters framework.

With this conditional state, we define the corresponding conditional probability density $P(\alpha|\beta_0)$ within the Klein--Gordon framework. We find, for all clock parameters $\beta_0$, that this  probability vanishes in the limit $\alpha \to -\infty$, corresponding to zero volume  of the universe.
This result indicates that the big bang singularity of the classical theory is effectively resolved in the Page-Wootters framework formulated within quantum cosmology. 

The requirement of a positive-definite conditional probability density imposes a constraint on the admissible values of the clock parameter $\beta_0$. In particular, the probability density can become negative in certain regions of the volume variable $\alpha$. To avoid such unphysical behavior, one must restrict $\beta_0$ to lie above a minimum value $\beta_{0,\min}$, which depends sensitively on the parameters ($k_0, \sigma$) of the underlying wavepacket. 

A key role in determining this bound is played by the mean value $k_0$ of the Gaussian distribution in momentum space. When $k_0$ is close to zero, the contributions from the left-moving and right-moving modes, ($e^{ik\alpha}$ and $e^{-ik\alpha}$), are almost symmetric and can interfere nearly destructively. If $\sigma \ll k_0$, fewer modes effectively contribute to the correlated structure of the wavepacket, which in turn imposes a stronger restriction on the lower bound of the clock parameter $\beta_0$. The possibility of achieving $\beta_{0,\min}$ close to zero arises only when the wavepacket is sufficiently broad, that is, when $\sigma \gg k_0$.

In contrast, when $k_0$ is sufficiently large, the wavepacket is peaked farther away from zero momentum, and a high asymmetry between left-moving and right-moving modes reduces the effectiveness of their mutual destruction. As a result, the interference effects responsible are weakened and a sufficiently large number of modes can participate in the correlated structure of the wave function. In this case, the lower bound on $\beta_0$ can approach close to zero for a wider range of parameters, including both $\sigma \gg k_0$ and $\sigma \ll k_0$. 

In summary, our analysis demonstrates that the Page-Wootters framework provides a viable approach to addressing foundational issues in quantum cosmology, particularly the problem of time and the resolution of the classical big bang singularity. The emergence of a nonsingular, well-defined probabilistic description highlights the potential of relational formulation in understanding the quantum nature of the early universe.

\section*{Acknowledgements}
Vishal is supported through a Research Fellowship from the Ministry of Human Resource Development (MHRD), Government of India. The Authors would like to thank the Indian Institute of Technology Guwahati for providing computing facilities.

\end{document}